# Ultra Low-Power All-Optical Switching

Marin Soljačić[(*)], Elefterios Lidorikis[(*)], J.D.Joannopoulos[(*)], and Lene Vestergaard Hau[(+)]

*(*) Physics Department, MIT, Cambridge, MA 02139*
*(+) Lyman Laboratory, Harvard University, Cambridge, MA 02138*

Using analytical modeling and detailed numerical simulations, we investigate properties of hybrid systems of Photonic Crystal micro-cavities which incorporate a highly non-linear Ultra Slow Light medium. We demonstrate that such systems, while being miniature in size (order wavelength), and integrable, could enable ultra-fast non-linear all-optical switching at single photon energy levels.

For many important applications (e.g. quantum information processing, integrated all-optical signal processing etc.) it would be highly beneficial to have strong and nearly instantaneous interaction of light with light, preferably happening in a minimal volume. This can be achieved, in principle, by exploiting intrinsic material non-linearities. Unfortunately, such non-linearities are well known to be extraordinarily weak, thus one is forced to make undesirable compromises on interaction time, device-length, and/or power. To optimize the effects, we combine two approaches to enhance optical non-linarities. One is structural: we design a structure whose geometrical properties enhance the non-linear interaction; due to their extraordinary opportunities for controlling the flow of light, photonic crystals (PhCs) [1,2,3], have been proven to be particularly suitable for this purpose [4,5,6,7]. The other approach is to use an Ultra Slow Light (USL) medium with extremely large non-linear optical response. Non-linear Kerr coefficients 12 orders of magnitude larger than in AlGaAs have been measured in such systems [8]. In this letter we demonstrate, through analytical theory and detailed numerical simulations, how combining the structural non-linearity enhancement offered by PhC cavities together with USL effects, can lead to all-optical switches of unprecedented characteristics; such switches can be less than $\lambda^3$ in size, with switching times faster than 100ps, and operating at single photon levels.

Let us begin with a simple argument of how such a device would operate. Imagine a resonant cavity with one input, and one output port, and narrow resonant transmission $(P_{OUT}(\omega)/P_{IN}(\omega))$ width $\Gamma=\omega_{RES}/2Q$, where $Q$ is the quality factor of the cavity, and $\omega_{RES}$ is the resonant frequency. In order to perform switching of an incoming probe field, one applies a stimulus to change the index of refraction $n$ inside the cavity, thereby shifting the resonance width by $\delta\omega_{RES}/\omega_{RES}\sim\delta n/n$. The switching requirement is: $\delta\omega_{RES}>\Gamma$. So, the larger the $Q$, the smaller the required $\delta n$ is. Imagine that $\delta n$ is induced through a non-linear Kerr effect by some other (control) field **E**: $\delta n \propto n_2|\mathbf{E}|^2$, where $n_2$ is the Kerr coefficient of the underlying material. If the incoming power of the control field is $P$, then the energy stored in the cavity $U=2QP/\omega_{RES}$, so $|\mathbf{E}|^2 \sim U/(\varepsilon V_{MODE}) \sim QP\omega_{RES}/(\varepsilon V_{MODE})$, where $V_{MODE} \equiv \dfrac{\int_{MODE} d^3x \varepsilon |\mathbf{E}|^2}{\varepsilon|\mathbf{E}_{MAX}|^2}$ is the modal volume of the cavity, and $\mathbf{E}_{MAX}$ is the peak electric field in the cavity. Combining all this, we conclude that the power needed to operate the device is $P \propto n_2 V_{MODE}/Q^2$; one $Q$-factor comes from the field-enhancement effects inside of the cavity, while the other $Q$-factor comes from decreased $\delta n$ requirements due to the narrower resonance width. It is important to emphasize that (in contrast to what typically happens in electronic switching schemes,) the energy in the fields is not actually absorbed, but rather re-directed during the switching processes: it does not cause heating of the system, and could even be reused. PhCs enable microcavities that can have small modal volumes and large $Q$s at the same time, thus providing optimal all-optical switching geometries. When implemented in AlGaAs, at $\lambda_{RES}=1.55\mu m$, and $Q=4000$, such switches can operate with $P\approx 5mW$ [6]. Therefore, if USL (whose $n_2$ can be 12 orders of magnitude larger than $n_2$ of AlGaAs [8]) is used as non-linear medium in such cavities instead of AlGaAs, power levels as low as $P\sim 10^{-15}W$ might be achievable. The point of this letter is to demonstrate that systems of the class we describe here are very natural for implementing extremely low power all-optical signal processing. To our knowledge, single-photon non-linear behavior of cavity-EIT (Electro-magnetically Induced Transparency) has only been discussed qualitatively using generic or heuristic models [9,10,11]. In this letter, we present results of realistic numerical experiments (including material and radiative losses) on an exemplary system of a PhC microcavity containing a single USL atom. In particular, we perform finite difference time domain (FDTD) simulations with perfectly matched layer (PML) boundary conditions [12], which simulate Maxwell's equations (including dispersion) for such a system exactly (apart for the discretization). Such simulations are known to be able to reproduce true experimental results very faithfully (hence the term "numerical experiments"), so the work presented here should act as a motivation to implement the systems we study in a true experimental setting.

For ease of fabrication, consider a PhC microcavity, as shown in Figure 1. This is a hybrid configuration, where the resonance is confined laterally by index-guiding and axially by the 1D PhC gap. For simplicity, we model a 2D system, since the essential physics is the same as that of its 3D counterpart, but numerical requirements are now much more tractable. The microcavity in Figure 1 has only a single resonance that is equally (and weakly) coupled to an input and output waveguide with:

$$T(\omega) \equiv \frac{P_{OUT}(\omega)}{P_{IN}(\omega)} = \left| \frac{i\Gamma_{IO}}{\omega - \omega_{RES} + i(\Gamma_{IO} + \Gamma_{RAD} + \Gamma_{ABS})} \right|^2, \quad (1)$$

where $P_{OUT}$ & $P_{IN}$ are outgoing & incoming powers, $\Gamma_{IO}$, $\Gamma_{RAD}$, and $\Gamma_{ABS}$ are respectively, the widths due to coupling to the waveguides, loss from the cavity due to the coupling to the free-space radiation modes, and the intrinsic material absorption, and $\omega_{RES}$ is the resonant frequency. We neglect the absorption of the host material (but not of the USL material) since it is typically small, and the presence of USL material does not increase its effects [17]. In that case, the transmission through the cavity is given by the dashed blue curve in Figure 2. If there were no radiation losses ($\Gamma_{RAD}=0$), this curve would peak at 100% transmission.

Consider now the presence of a single USL atom [13] at the center of the microcavity. This could be implemented by using AFM techniques, solid-state USL materials [14], or a single-gas-atom PhC microcavity [15]. The relevant atomic levels of such an atom are shown in Figure 3A. In general, one would need ensure that each of the relevant atomic transitions coincides with an *even* resonant mode of the cavity. (E.g. for the particular system of Figure 1, the resonance is even and all transitions would have to fit within its single resonance width, which is $\approx \omega_{RES}/692$). Next, one would introduce a coupling field at frequency $\omega_{23}$ into the cavity, in order to establish USL for the probe frequencies $\omega$ which are close to the $\omega_{13}$ transition. The polarizability of a typical USL atom is shown in Figure 3B.

Introduction of a highly dispersive polarizable object into a cavity has two important effects. First, it changes the resonant frequency of the cavity. According to perturbation theory [6]:

$$\widetilde{\omega}_{RES} \approx \omega_{RES}\left[1 - \frac{\alpha}{2\varepsilon V_{MODE}}\right], \qquad (2)$$

where the induced dipole moment $\mathbf{p}=\alpha\mathbf{E}$ (here, $\mathbf{E}$ is the electric field at the position of the dipole, and $\alpha$ is the atomic polarizability), $V_{MODE}$ is the modal volume, and $\varepsilon$ is the dielectric constant of the host medium. Note that since $\alpha$ is in general complex, Eq.(2) also causes an effective change in $\Gamma_{ABS}$ as: $\widetilde{\Gamma}_{ABS} \approx \Gamma_{ABS} + \frac{\omega_{RES}\,\mathrm{Im}\{\alpha\}}{2\varepsilon V_{MODE}}$; for most USL systems of interest this effect will be barely noticeable, but we include it here for completeness. Second, this object results in a change of geometry of the cavity, thereby modifying its coupling to the free-space radiation modes ($\Gamma_{RAD}$). For most cavities, the change in $\Gamma_{RAD}$ will also be unnoticeable. Still, it can be understood as follows. Usually, power scattered by an induced dipole is $\propto|\mathbf{p}|^2$. However, in our case, both the induced dipole, and the cavity mode itself scatter out of the cavity a significant portion of power into the same single mode (dipole far-field radiation expansion). Consequently, their fields (rather than powers) add, and the change in the radiated power $\Delta P_{RAD}$ has a component linear in $\mathbf{p}$. Since $Re\{\alpha\} \gg Im\{\alpha\}$ for a typical USL application, we can approximate:

$\Gamma_{RAD} \approx \Gamma_{RAD}(\mathbf{p}=0) + \xi Re\{\alpha\} + \ldots$ (3)

where, $\xi$ is determined by the geometry of the cavity, and has to be calculated for each cavity separately: one simulates systems with a few different values of $\alpha$, and fits $\Gamma_{RAD}$ to a straight line. For our 2D cavity from Figure 1, we calculate $\xi \approx 0.0012c/(a^3\varepsilon_0)$.

The enormous dispersive behavior [8] like the one shown in Figure 3B drastically narrows the transmission resonance width of the cavity for probe frequencies $\omega$ close to $\omega_{13}$. Intuitively, the large dispersion implies low group velocity, so each "bounce" between the two mirrors of the cavity takes longer time, meaning that the light spends longer time in the cavity. For the particular case of the dipole shown in Figure 3B, the FDTD calculation of narrowing gives a factor of $\approx 3.3$ [16], as shown by the solid blue line in Figure 2. Despite the fact that the light spends much more time in the cavity, the coupling to the free-space radiation modes is not increased [17], so the peak transmission is unchanged. A way to see that this has to be so is to note that since $\omega_{13}$ coincides with $\omega_{RES}$, the induced dipole moment is zero at $\omega_{RES}$, so the system behavior (and hence its peak transmission) is unchanged at that particular frequency. The behavior of the system can also be well described with our analytical model, as shown by the black dashed line coinciding with the solid-blue curve in Figure 2.

Finally, we introduce an additional third (control) field into the cavity, with a frequency close to $\omega_{24}$, in order to perform the switching of the probe field $\omega$ [18,19,20]. Presence of this control field causes a Stark shift of level $|2\rangle$, thereby sliding the whole dispersion curve in Figure 3B sideways. This switching behavior is displayed by red, green, and magenta curves in Figure 2; the corresponding dashed black lines represent the results of the analytical modeling of the system. Clearly, perturbation theory models the true behavior very faithfully.

We can now use the analytical model to understand the behavior of such devices in various USL parameter regimes. We start by writing the expression for $\alpha$ of an USL atom, for $\omega$ close to $\omega_{13}$ using arguments similar to those of Refs. 21 and 22:

$$\alpha \approx \frac{6e^2 f_{13}}{m_e \omega_{13}} \left[ \frac{\Delta_P}{|\Omega_C|^2} + 2i\Gamma_3 \left( \frac{\Delta_P}{|\Omega_C|^2} \right)^2 \right], \qquad (4)$$

where $f_{13}$ is the oscillator strength of $|3\rangle \rightarrow |1\rangle$ transition, $\Gamma_3$ is the decay width of state $|3\rangle$ (which can in general be different than the free-space decay width of state $|3\rangle$: in our case it is larger by a factor $\sim Q\lambda^3/V_{MODE}$ due to cavity-QED effects), $\Omega_C$ is the Rabi frequency of the coupling field (at frequency $\omega_{23}$), and:

$$\Delta_P \equiv \omega - \left(\omega_{13} - \frac{|\Omega_{24}|^2}{4\Delta\widetilde{\omega}_{24}}\right), \tag{5}$$

where $\Omega_{24}$ is the Rabi frequency of the control field, $\Delta\widetilde{\omega}_{24} = \Delta\omega_{24} - i\gamma_{24}$, $\Delta\omega_{24}$ is the difference in frequencies between the control field, and $\omega_{24}$, while $\gamma_{24}$ is the decay width of the $|4\rangle \rightarrow |2\rangle$ transition. For the application of interest, we can approximate: $\Delta\widetilde{\omega}_{24} \approx \Delta\omega_{24}$. Next, we substitute Eqs.(2),(3)&(4), into Eq.(1) to obtain:

$$T(\omega) = \left| \frac{i\Gamma_{IO}}{\omega - \omega_{RES} + \frac{c}{v_G}\Delta_P + i\left[\Gamma_{RAD}(\mathbf{p}=0) + \xi\frac{6e^2 f_{13}}{m_e\omega_{13}}\frac{\Delta_P}{|\Omega_C|^2} + \Gamma_{IO} + \frac{6e^2 f_{13}\Gamma_3}{V_{MODE}\varepsilon m_e}\left(\frac{\Delta_P}{|\Omega_C|^2}\right)^2\right]} \right|^2, \tag{6}$$

where we define: $\dfrac{3e^2 f_{13}}{m_e V_{MODE}\varepsilon}\dfrac{1}{|\Omega_C|^2} \equiv \dfrac{c}{v_G}$; $v_G$ has a simple physical interpretation: it is the group velocity of propagation in uniform USL media, consisting of (same) USL atoms, but with atomic density $1/V_{MODE}$.

In regimes of interest, $c/v_G \gg 1$, so the real part of the denominator of Eq.(6) can be approximated as $\omega - \omega_{RES} + \dfrac{c}{v_G}\Delta_P \approx \dfrac{c}{v_G}\Delta_P$, so $T(\Delta_P=0) \approx \Gamma_{IO}^2/[\Gamma_{IO}+\Gamma_{RAD}(\mathbf{p}=0)]^2$, which is the same as the peak transmission of the cavity without the USL atom. Thus, despite the extreme width-narrowing, and irrespective of the $\Omega_{24}$-induced resonance shifts, the peak transmission is always the same. Consequently, the somewhat worrisome effect (apparent by the green curve in Figure 2) of reduced transmission during the switching operation disappears as one operates in the regime of strong USL effects ($c/v_G \gg 1$). Furthermore, for properly designed microcavities, and properly implemented USLs, absorption (term proportional to $\Gamma_3$ in Eq.(6)), and changes to the cavity geometry (term proportional to $\xi$ in Eq.(6)) can both be neglected. Thus, the width of the transmission curve (Eq.(6)) is given by $\approx[\Gamma_{IO}+\Gamma_{RAD}(\mathbf{p}=0)](v_G/c)$, so the narrowing factor is $\approx v_G/c$. To obtain switching, we need to shift the resonance by more than its width:

$$\frac{c}{v_G}\frac{|\Omega_{24}|^2}{4|\Delta\widetilde{\omega}_{24}|} > \Gamma_{RAD}(\mathbf{p}=0)+\Gamma_{IO}. \qquad (7)$$

The optimal efficiency of our systems is now apparent from Eq.(7). The right hand side of the equation is the transmission width of the cavity without the USL atom present: the larger its $Q$, the more efficient the system is. The left hand side is just the Kerr-effect induced change in the resonant frequency of the cavity. The strength of this Kerr-effect is greatly enhanced because of three factors: $(c/v_G)$ can be made large, $\Delta\widetilde{\omega}_{24}$ can be made small (so we are exploring non-linearities close to the resonance which one cannot do in usual non-linear systems because of huge absorption), and for a given incoming power $P_{24}$, the cavity enhancement effects and the small modal volume both make $\Omega_{24}$ large.

Before concluding, we estimate quantitative performance characteristics of a 3D device of the type we describe. First, we assume that the modal extent in the direction out of the page in Figure 1 is roughly the same as the modal extent in the direction perpendicular to the waveguide in the plane of the figure. This gives an estimate of $V_{MODE}\approx 0.009\lambda_{RES}^3$. As an example, we will use a resonance of the sodium atom with $\lambda_{RES}=589nm$. We assume resonance-narrowing factor due to USL of $c/v_G\approx 30$, leading to a transmission width (and hence the available operational bandwidth in $\omega$) $\approx 25GHz$. (For comparison, if we chose to use the experimental parameters of Ref.8, the narrowing factor would be $>10^7$!) To implement switching, the induced Stark shift is: $|\Omega_{24}|^2/|\Delta\omega_{24}|>25GHz$ [13]. So, if we take $\Delta\omega_{24}=60GHz$ (which would provide us with ~10GHz operational bandwidth for the control field), the needed intensity of the control field in air (for sodium) would be $I_{24}\approx 50GW/m^2$ [8], while the field inside the cavity is: $|\mathbf{E}_{24}|^2=2I_{24}/(c\varepsilon_0)$. The needed input power $P_{24}=\omega_{RES}U_{24}/(2Q_{24})=(\varepsilon\pi V_{MODE}I_{24})/(2\varepsilon_0\lambda_{RES}Q_{24})$, where $U_{24}$ is the control field's modal energy, and $Q_{24}\approx 692$ is its transmission $Q$ for the cavity of Figure 1. We conclude that $P_{24} \approx 4.3 \mu W$. With similar reasoning, we can show that the power in the coupling field needs to be $P_C \approx 10 \mu W$. Finally, the number of the control-field photons needed to be present in the cavity in order to induce the switching is $N_{24} = \dfrac{V_{MODE} \varepsilon |\mathbf{E}_{24}|^2}{2\hbar \omega_{24}} \approx 11$. Each of these photons spends ~2ps in the cavity, while the switching time is ~100ps; so the switching is performed by a total of $N_{24}$~500 photons. By exploring even more extreme regimes of USL parameters and/or higher $Q$ PhC cavities ($Q=45000$ PhC micro-cavity has been demonstrated experimentally recently [23]), one can easily reach the single-photon optical non-linearity operation regime, which has been elusive thus far.

In conclusion, we describe a class of microcavity devices with extraordinary optical non-linear properties. Of course, physical implementation of such systems will entail overcoming many technical hurdles (e.g. identifying an optimal atomic system, matched to the desired operational wavelength, etc.) As noted above, extremely low power levels required for optical non-linearities with the proposed hybrid system would allow exploration of an exciting new physics regime where, e.g. quantum fluctuations in the fields should be an important non-linear effect. Moreover, such devices could prove useful for all-optical quantum information processing. Finally, it would be interesting to explore the possibility that USL could be automatically established in some solid-state single-atom systems, even if one does not apply the coupling field: the electric field of the host material could (under proper conditions) play the role of the coupling field.

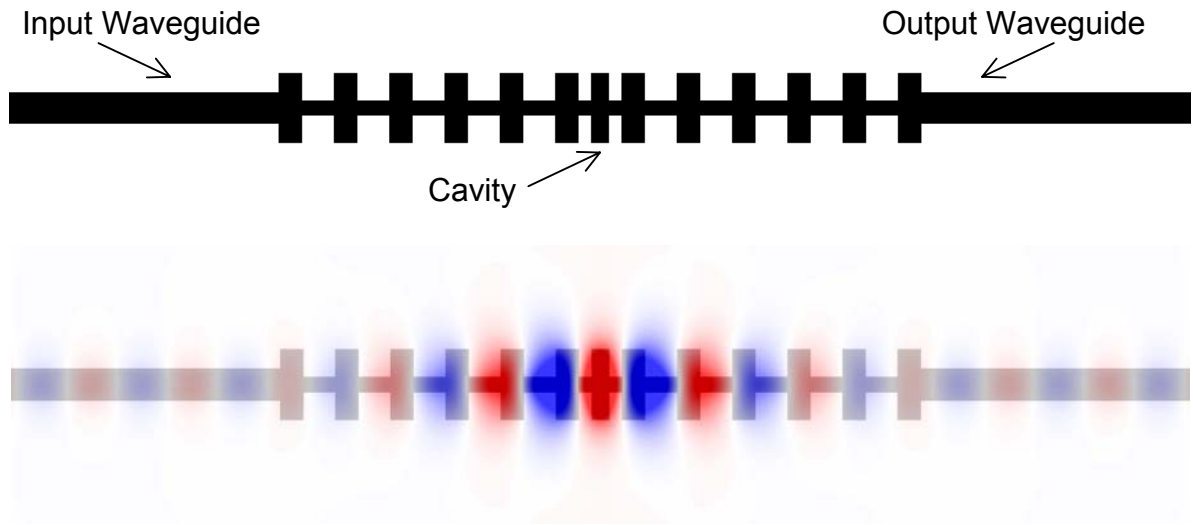

**Figure 1:** PhC microcavity studied in this letter (dielectric profile shown on the top), and the electric field (all pointing out of the plane) of its resonant mode (shown on the bottom, together with the high index material (gray)). High index material has $\varepsilon=12$, and is surrounded with air ($\varepsilon=1$). The cavity is implemented by introducing a defect into a periodic structure, of period *a*. Each periodic cell consists of a thick segment (thickness *1.25a*, length *0.4a*), followed by a thin segment (thickness *0.25a*, length *0.6a*). The defect is introduced by narrowing the length of the central thick element to *0.3a*, and narrowing the length of its two neighboring thin elements to *0.25a*. The incoming and outgoing waveguides have thickness *0.55a*. The runs are performed at a numerical grid-resolution of *40pts/a*. Consistency is checked at *20pts/a*, and *80pts/a*.

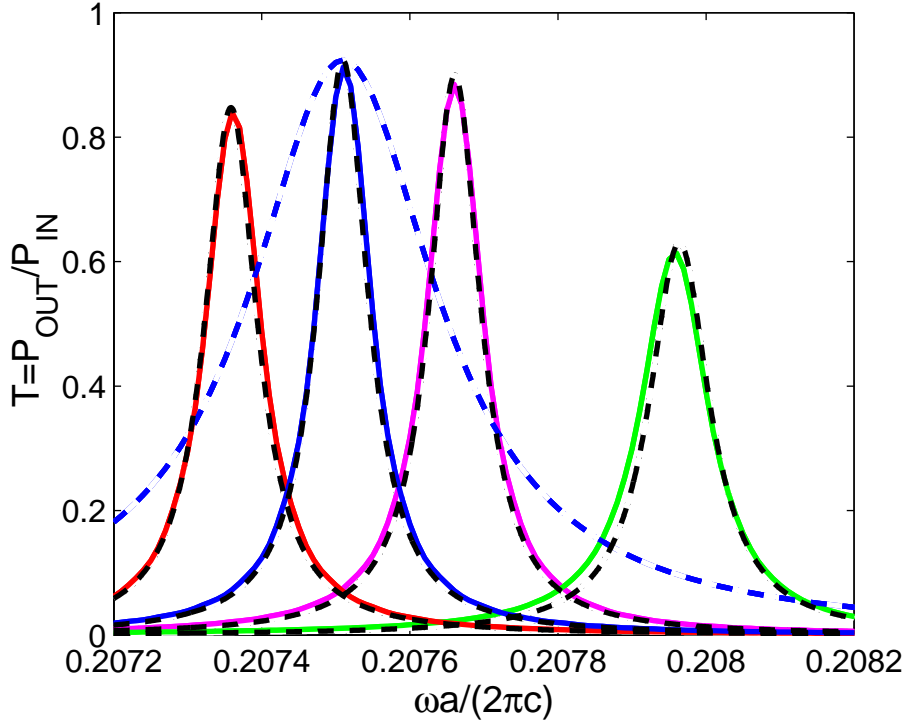

**Figure 2:** Transmission through the system of Figure 1. The dashed blue curve presents the FDTD calculation when the USL atom is not present. The solid blue curve presents the FDTD calculation with the USL atom present (dispersion given by Figure 3B), and $\omega_{13}$ exactly coinciding with the resonance of the cavity without the USL atom present. Red, green, and magenta curves present FDTD calculations when dispersion in Figure 3B is shifted sideways as: red ($\omega_{13} \rightarrow \omega_{13}*0.999$), magenta ($\omega_{13} \rightarrow \omega_{13}*1.001$), and green ($\omega_{13} \rightarrow \omega_{13}*1.003$). The dashed black curves are predictions of the perturbation theory for their corresponding curves: they are obtained as follows. First, we assume linear dependence of $Re\{\alpha\}$ close to $\omega_{13}$, and quadratic dependence of $Im\{\alpha\}$ close to $\omega_{13}$, with fit parameters obtained from Figure 3B; these are needed for Eq.(2). Second, with a series of independent FDTD calculations we obtain a linear fit to $\Gamma_{RAD}$ in $Re\{\alpha\}$, as required by Eq.(3). Next, we obtain $\Gamma_{IO}$, and $\omega_{RES}$ from the dashed blue curve above, and calculate $V_{MODE}$ with an independent simulation. Finally, we substitute the expressions obtained by Eqs.(2)&(3) in this manner into Eq.(1), in order to obtain the black dashed curves shown here.

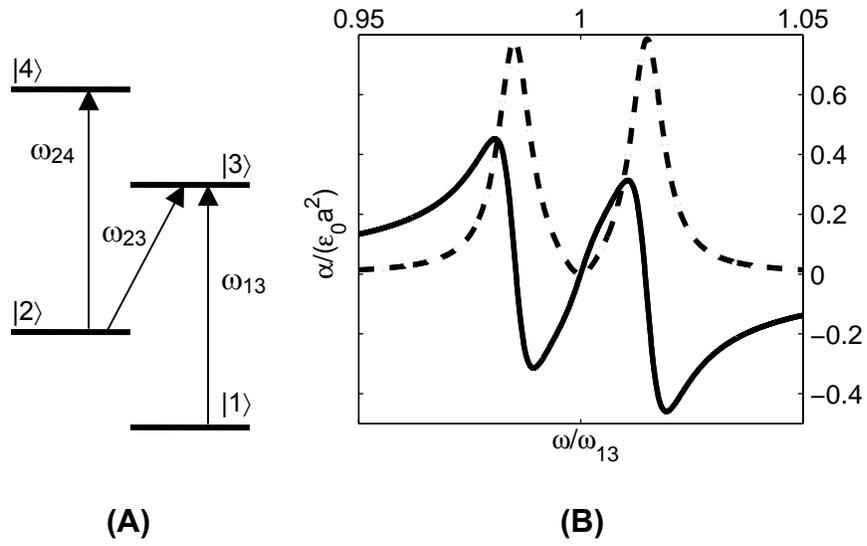

**Figure 3: (A)** Schematic of atomic levels in a typical USL system. **(B)** Normalized polarizability of the USL atom of interest: solid line is *Re{α}*, and dashed line is *Im{α}*.

**REFERENCES:**

1   E. Yablonovitch, Phys. Rev. Lett. **58**, 2059 (1987).

2   S. John, Phys. Rev. Lett. **58**, 2486 (1987).

3   J. D. Joannopoulos, R. D. Meade, and J. N. Winn, *Photonic Crystals: Molding the flow of light* (Princeton University Press, Princeton, N.J., 1995).

4   A.Yariv, Y.Xu, R.K.Lee, and A. Scherer. Opt. Lett. **24**, 711 (1999).

5   Marin Soljacic, S.Johnson, S.Fan, M.Ibanescu, E.Ippen, and J.D.Joannopoulos. JOSA B **19**, 2052, (2002).

6   Marin Soljacic, Mihai Ibanescu, Steven G. Johnson, Yoel Fink, and J.D.Joannopoulos. Phys. Rev. E **66**, 055601(R), (2002).

7   S.F. Mingaleev, and Y.S. Kivshar, JOSA B **19**, 2241 (2002).

8   L.V.Hau, S.E.Harris, Z.Dutton, and C.H.Behroozi, Nature, **397**, 594 (1999).

9   S. Rebic, S. M. Tan, A. S. Parkins and D. F. Walls, J. Opt. B: Quantum Semiclass. Opt. **1**, 490-495 (1999).

10  M.J.Werner and A.Imamoglu, Phys. Rev. A, **61**, 011801(R), (1999).

11  S. Rebic, S. M. Tan, A. S. Parkins and D. F. Walls, "Photon Blockade with a Single Atom", preprint

12  For a review, see A. Taflove, *Computational Electrodynamics: The Finite-Difference Time-Domain Method* (Artech House, Norwood, Mass., 1995).

13  Since there is only one atom, the Doppler broadening in gaseous USL systems, or the inhomogeneous broadening in solid-state systems (e.g. 5GHz in Pr:YSO [14]) due to the host is not an issue anymore, while the remaining uncertainty in the exact level positions (because of the influence of the host, or because of Doppler broadening) can be much smaller than the cavity resonance width.

14  A.V.Turukhin, V.S.Sudarshanam, M.S.Shahriar, J.A.Musser, B.S.Ham, and P.R.Hemmer, Phys. Rev. Lett. **88**, 023602, (2002).

15  J.Vuckovic, M.Loncar, H.Mabuchi, and A.Scherer, Phys. Rev. E, **65**, 016608, (2001).

16  In our numerics, we model the USL atom as a small-area object ($2.4*10^{-4}\lambda_{RES}^2$), with large, highly-dispersive susceptibility. The required dispersion shape is obtained with two absorption lines sandwiching a gain line; such a shape closely resembles a typical USL dispersion.

17  "Enhancement of cavity lifetimes using highly dispersive materials", Marin Soljacic, Elefterios Lidorikis, Lene Vestergaard Hau, J.D. Joannopoulos, submitted for publication.

18  H.Schmidt, and A.Imamoglu, Optics Letters, **21**, 1936 (1996).

19  S.E.Harris, and L.Hau, Phys. Rev. Lett. **82**, 4611, (1999).

20   Danielle A. Braje, Vlatko Balic, G. Y. Yin, and S. E. Harris, Phys. Rev. A **68**, 041801(R), (2003).

21   S.E.Harris, J.E.Field, and A.Kasapi, Phys Rev A, **46**, R29, (1992).

22   Zachary Dutton, *Ultra-slow, stopped, and compressed light in Bose-Einstein condensates*, Harvard University doctoral dissertation, (2002).

23   Yoshihiro Akahane, Takashi Asano, Bong-Shik Song, and Susumu Noda, Nature **425**, 944, (2003).